\def\today{\ifcase\month\or
  January\or February\or March\or April\or May\or June\or
  July\or August\or September\or October\or November\or December\fi
  \space\number\day, \number\year}

\documentstyle[preprint,aps]{revtex}

\begin{document}
\draft
\title{Anderson Model in a Superconductor:\\
$\Phi $-Derivable Theory}
\author{Ari T.\ Alastalo$^{\rm (a)}$, Robert J.\ Joynt$^{\rm (b)}$ 
and Martti M.\ Salomaa$^{\rm (a,b)}$}
\address{$^{\rm (a)}$Materials Physics Laboratory,
P.O.\ Box 2200 (Technical Physics)\\
Helsinki University of Technology, FIN-02015 HUT, Finland}
\address{$^{\rm (b)}$Department of Physics and Applied Superconductivity Center\\
University of Wisconsin-Madison, Madison, WI 53706, USA}
\date{\today}

\maketitle

\newpage
\begin{abstract}
We introduce a new $\Phi $-derivable approach for the Anderson 
impurity model in a BCS superconductor. The regime of validity of 
this conserving theory extends well beyond that of the Hartree-Fock 
approximation. This is the first generalization of the $U$-perturbation 
theory to encompass a superconductor.
\end{abstract}

\vfill
\pacs{PACS Nos: 72.15.Qm, 75.20.Hr, 75.30.Mb, and 74.25.-q}

\narrowtext

The Anderson impurity model [1] provides one of the most versatile 
field-theoretic descriptions for interacting correlated electrons within 
condensed-matter physics [2]. The Anderson hamiltonian in a normal 
metal is 
\begin{equation}
{\cal H}_A = {\cal H}_s + {\cal H}_{sd} + {\cal H}_d + {\cal H}_U \, ,
\end{equation}
where ${\cal H}_s = \sum _{k,\sigma} \varepsilon _{k \sigma}n_{k \sigma}$
describes the electron gas, ${\cal H}_{sd}=\sum _{k, \sigma}
(V_kc^{\dagger }_{k \sigma}d_{\sigma} + V_k^{\ast }d_{\sigma}^{\dagger }
c_{k \sigma})$ is the admixture interaction, ${\cal H}_d = \sum _{\sigma}
E_{\sigma }n_{\sigma }$ represents the $d$-electron level and ${\cal H}_U = 
Un_{\uparrow}n_{\downarrow}$ denotes the Coulomb-repulsion interaction. 
This model describes 
the continuous transition of a nonmagnetic resonant level (for $U\ll \Gamma$, 
where $\Gamma = \pi N(0) \langle \vert V\vert ^2\rangle$) 
to a magnetic atom ($U\gg \Gamma$), and one can consider 
treating either $V$ or $U$ perturbatively. For $\Gamma /U\ll 1$, 
in the magnetic Schrieffer-Wolff (SW) limit [3], the Anderson hamiltonian reduces to 
the $s-d$ hamiltonian, {\it i.e.}, the Kondo model. This limit 
and the magnetic-nonmagnetic transition have been succesfully
treated within the renormalization group (RNG) program [4]. However,
simpler controlled approaches would be highly desirable due to the 
wide applicability of the Anderson model and its variants to many physical
systems of interest.   

After its initial introduction to describe the nonmagnetic-magnetic 
transition of impurities in otherwise nonmagnetic metals, and the 
associated many-body Kondo phenomenology, the Anderson model 
has been extensively applied and generalized to also describe 
interacting pairs of impurity atoms
in metals (the Alexander-Anderson model), valence fluctuations and 
heavy-fermion materials (the periodic Anderson model), 
chemisorption (the Anderson-Newns model) and charging phenomena 
and Coulomb blockade in quantum dots and quantum-dot arrays.
It is also of great inherent interest to consider the Anderson model 
in a superconductor and the appropriate theoretical approaches to 
this problem. In particular, the RNG approach has not been 
generalized to this case and the Bethe-Ansatz (BA) method fails
to be suitable since the superconducting electronic spectrum does 
not fulfill the requirement of a linear ($\varepsilon _k \propto 
k $) dispersion relationship, necessary for the applicability of 
the $k$-state enumeration within the BA scheme. 

Yosida and Yamada [5] first pointed out that 
in order to obtain the single-particle Green's function and 
the local density of impurity $d$-states, it is 
useful to study the many-body perturbation theory with respect 
to $U$. Thus one considers the Anderson hamiltonian as
\begin{equation}
{\cal H}_A= {\cal H}^0_{HF} - U\langle n_{\uparrow}\rangle 
\langle n_{\downarrow }\rangle + {\cal H}^{\prime }_U \, ,
\end{equation}
where  ${\cal H}^0_{HF} - U\langle n_{\uparrow}\rangle 
\langle n_{\downarrow }\rangle $ is up to the constant energy shift, 
$U\langle n_{\uparrow}\rangle \langle n_{\downarrow }\rangle $,
the unperturbed Hartree-Fock (HF) hamiltonian and 
\begin{equation}
{\cal H}^{\prime }_U = U \delta n_{\uparrow} \delta n_{\downarrow}
= U(n_{\uparrow } - \langle n_{\uparrow }\rangle )(n_{\downarrow } - \langle n_{\downarrow }\rangle )
\end{equation}
is the Coulomb-repulsion term, treated as the perturbation, see Fig.~1. 
Yosida and Yamada, in their original paper [5], used a complicated 
formalism utilizing Pfaffian determinants to derive the $U$-perturbation
theory; Yamada [6] first presented  the numerically evaluated $d$-electron 
spectral density function using the $U^2$ selfenergy in 
the low-temperature limit.
This approach may easily be generalized to arbitrary temperatures 
in normal metals [7],
for which one may introduce the $d$-electron impurity selfenergy 
$\Sigma _N(\omega )$ through
\begin{equation}
G_N (\omega ) = \Bigl ((G_{HF})^{-1} - \Sigma _N(\omega )\Bigr )^{-1} \, .
\end{equation}
Here the HF propagator is
\begin{equation}
G_{HF}(\omega ) = - \Bigl (\omega - E_{HF}  - F(\omega )\Bigr )^{-1} \; ,
\end{equation}
with $F(\omega )=\sum _k \vert V_k\vert ^2G^0_k(\omega ) \approx 
- i\Gamma$ and $G^0_k(\omega ) = (\omega -\varepsilon _k)^{-1}$. Above, 
$E_{HF} = U + \langle n\rangle $ is the Hartree-Fock energy. 
Here and in what follows we omit the spin indices for brevity; hence our
expressions are valid for zero field.
Using the second-order free-energy functional $\Phi _N^{(2)}$ in Fig.~2, 
one easily finds that the $U^2$ selfenergy may be obtained by cutting
the $d$-electron propagator line in the diagram as 
\begin{equation}
\Sigma _N= \delta \Phi _N/\delta G _N \, . 
\end{equation}
Consequently, the imaginary part of the second-order contribution (in $U$) to the
impurity self-energy is given by
\begin{equation}
\Sigma _N^{\prime \prime }(\omega ) = U^2
\int {{d\omega _1}\over {\pi}}\int {{d\omega _2}\over {\pi}}\int d\omega _3 \, 
\delta (\omega - \omega _1 -\omega _2 - \omega _3) 
F(\omega _1, \omega _2, \omega _3) 
G_{HF}^{\prime \prime }(\omega _1)
G_{HF}^{\prime \prime }(\omega _2) 
G_{HF}^{\prime \prime }(\omega _3) \, .
\end{equation}
Here $F(\omega _1, \omega _2, \omega _3)$ abbreviates the 
following collection of thermal occupancy factors 
\begin{equation}
F(\omega _1, \omega _2, \omega _3) = 
\lbrack 1-f(\omega _1)\rbrack \lbrack 1-f(\omega _2)\rbrack 
\lbrack 1-f(\omega _3)\rbrack +  f(\omega _1)f(\omega _2)f(\omega _3) \, , 
\end{equation}
with $f(\omega )= (e^{\omega /T} + 1)^{-1}$ denoting the Fermi distribution function.

Note that the $U$-perturbation theory can be derived from 
a free-energy functional, $\Phi$. Therefore, this is a 
"conserving approximation" [8] for the many-body system, 
with positive definite spectral functions and with sum rules
fulfilled by construction.
The linear term in $U$, see Fig.~2, is the HF term, describing the motion of a 
$d$-electron with spin $\sigma $ in the mean field produced
by the $d$-electron with spin $-\sigma $. This mean field, 
or the expectation value $\langle n\rangle $, must be
computed selfconsistently. The  
quadratic term in $U$ describes the interaction, at the 
impurity site, of the localized spin fluctuations (LSF) represented 
by the particle-hole spin-susceptibility bubbles $(G_N\bar G_N)$.  

Yamada [6] first showed that the $U^2$-selfenergy 
$\Sigma ^{(2)}_N(\omega )$ yields a triple-peaked structure for the spectral 
density of the impurity atom. The sharp central peak for $T=0$ obtains the unitary limiting value at $\omega =0$. This approach was later 
generalized to arbitrary finite temperatures [7] and it was 
shown that the central zero-frequency 
peak has a sensitive $T$ dependence. 

The Anderson model in a superconductor is given by  
\begin{equation}
{\cal H}_{A,BCS} = {\cal H}_{BCS}+ {\cal H}_{sd} + {\cal H}_d + {\cal H}_U \, ,
\end{equation}
where the BCS hamiltonian is 
\begin{equation}
{\cal H}_{BSC} = \sum _{k,\sigma }\varepsilon _{k\sigma}n_{k\sigma}
-\sum _k (\Delta c^{\dagger }_{k\uparrow}c^{\dagger }_{-k\downarrow}
+ \Delta ^{\ast }c_{-k\downarrow }c_{k\uparrow }) \, .
\end{equation}
This model has been discussed in the HF [9, 10] and Schrieffer-Wolff [11]
limits.  In the HF approximation for a superconductor one truncates the 
Coulomb-interaction term as follows:
\begin{equation}
Un_{\uparrow}n_{\downarrow} \rightarrow 
U\langle n_{\uparrow}\rangle n_{\downarrow}
+ U\langle n_{\downarrow }\rangle n_{\uparrow }
+ U\langle d^{\dagger }_{\uparrow }d^{\dagger }_{\downarrow}\rangle 
d_{\downarrow }d_{\uparrow }
+ U\langle d_{\downarrow }d_{\uparrow}\rangle 
d^{\dagger }_{\uparrow }d^{\dagger }_{\downarrow } \, .
\end{equation}
Here the anomalous average 
$\langle d_{\downarrow }d_{\uparrow}\rangle $ 
induced at the impurity site presents another mean field, 
in addition to $\langle n\rangle $, which is to be
computed selfconsistently. 
The HF approximation to the Anderson model in a superconductor [9]
shares the same instability problem as the HF approximation in the
normal metal [1]: a spontaneous unphysical breaking of symmetry at
the impurity site. Our aim is to develop an approach which is free
from this HF instability and which enables one to go beyond the 
HF picture. In particular, we are interested to investigate the 
qualitatively new physical features, especially in the $d$-electron
density of states, due to increasing electron correlations
as $U$ increases beyond $\pi \Gamma $, where the HF solution
no longer provides quantitatively meaningful answers.

A generalization of the $U$-perturbation theory for the Anderson 
model in a superconductor has thus far not been discussed in the literature. 
The purpose of this Letter is to suggest
a new, conserving, self-energy $U$-perturbation expansion that is valid 
in a superconductor.  We note that the RNG approach has recently been 
generalized to treat magnetic impurities in superconductors, but thus far only 
for the $s-d$ model [12].

The matrix selfenergy expansion in the Nambu space is introduced as 
\begin{equation}
\hat G_S(\omega ) = \Bigl (({\hat G}_{HF}(\omega ))^{-1} 
- \hat {\Sigma}_S(\omega)\Bigr )^{-1} \, ,
\end{equation}
where the hat denotes matrices in the particle-hole Nambu space and 
the $d$-electron propagator in the HF approximation is given as
\begin{equation}
{\hat G}_{HF}(\omega ) = - \pmatrix{\omega -E_{HF} -F_{11}(\omega)&
U\langle d_{\sigma}d_{-\sigma}\rangle  - F_{12}(\omega) \cr
U\langle d_{-\sigma}^{\dagger}d_{\sigma }^{\dagger }\rangle - F_{21}(\omega )& 
\omega +E_{HF} - F_{22}(\omega )} ^{-1}\, ,
\end{equation}
which, for brevity, we denote here as: 
\begin{equation}
{\hat G}_{HF} (\omega ) = 
\pmatrix{{\cal G}(\omega )&{\cal F}(\omega ) \cr
{\bar {\cal F}}(\omega )& {\bar {\cal G}}(\omega )} \, .
\end{equation}
These propagators are illustrated graphically in Fig.\ 3 as lines 
with two arrows.

The energy-integrated Green's function (or generalized density of states) 
${\hat F}_S(\omega )$ in Eq.~(13) is 
\begin{equation}
{\hat F}_S(\omega ) = 
\pmatrix{F_{11}(\omega )&F_{12}(\omega ) \cr
F_{21}(\omega )& F_{22}(\omega )} 
= \sum _k \hat V_k^{\ast } \hat G^0_k(\omega ) \hat V_k\, ,
\end{equation}
where 
\begin{equation}
\hat V_k = \pmatrix{V_k & 0 \cr  0 & -V_k^{\ast }} \, .
\end{equation}
Above, in Eq.~(15), $\hat G^0_k(\omega )$ is the unperturbed Green's function
for the bulk superconductor:
\begin{equation}
\hat G^0_k(\omega ) = \pmatrix{\omega -\varepsilon _k&\Delta \cr
\Delta ^{\ast }& \omega +\varepsilon _{-k}}^{-1} \, . 
\end{equation}
The selfenergy in Eq.~(12) is a matrix in the Nambu space  
\begin{equation}
{\hat \Sigma}_S (\omega ) = 
\pmatrix{\Sigma _{11}(\omega )&\Sigma _{12}(\omega ) \cr
\Sigma _{21}(\omega )& \Sigma _{22}(\omega )} \, ,
\end{equation}
which may be obtained as
\begin{equation}
\Sigma _{ij} = \delta \Phi _S/\delta (G_S)_{ij} \, , 
\end{equation}
where $\Phi _S$ now denotes the free energy in a superconductor. 

The second-order free-energy term in $U$ for a superconductor, 
$\Phi _S^{(2)}$, is illustrated in Fig.~4 from which we obtain the following 
expression, accurate to $U^2$:
\begin{eqnarray}
\pmatrix{\Sigma _{11}^{\prime\prime}(\omega )&\Sigma _{12}^{\prime\prime}(\omega ) \cr
\Sigma _{21}^{\prime\prime}(\omega )& \Sigma _{22}^{\prime\prime}(\omega )} & = &
U^2 \int {{d\omega _1}\over {\pi}}\int {{d\omega _2}\over {\pi}}\int d\omega _3
\, \delta (\omega - \omega _1 -\omega _2 - \omega _3) 
F(\omega _1, \omega _2, \omega _3) \nonumber \\
& & \times \pmatrix{{\cal G}^{\prime \prime }(\omega _1)&
- {\cal F}^{\prime \prime }(\omega _1) \cr
- {\bar {\cal F}}^{\prime \prime }(\omega _1)& 
{\bar {\cal G}}^{\prime \prime }(\omega _1)} 
\Bigl ({\cal G}^{\prime \prime }(\omega _2){\bar {\cal G}}^{\prime \prime }(\omega _3)
- {\cal F}^{\prime \prime }(\omega _2) {\bar {\cal F}}^{\prime \prime }(\omega _3)\Bigr )\, ,
\end{eqnarray} 
where $F(\omega _1, \omega _2, \omega _3)$ is, again, given by Eq.~(8).
Note that the propagators ${\cal G}$, ${\bar {\cal G}}$, ${\cal F}$, 
and ${\bar {\cal F}}$ in the above expression may be evaluated in the 
HF approximation, see Eq.~(13), which already contains the pairing 
ineraction ${\cal H}_{BCS}$ to infinite order in the unperturbed 
hamiltonian ${\cal H}^0$. Therefore, the $\omega $ integrals in 
Eq.~(20) are rather complicated: they contain delta-function 
contributions from the bound states and also continuum 
contributions. Trivially, one observes that the normal-state limit, Eq.~(7), is 
obtained consistently from Eq.~(20) when $\Delta \rightarrow 0$ 
and that the diagrams in Fig.~4 reduce to those in Fig.~2 for
$\Delta \rightarrow 0$. 

The free-energy diagrams in Fig.~4 now comprise of 
two contributions linear in $U$, corresponding to the selfconsistent
occupation-number field $\langle n\rangle $, and the induced 
anomalous average (the proximity pairing at the impurity $d$-orbital site) 
$\langle d_{\downarrow }d_{\uparrow }\rangle $. Furthermore, 
there now occur three terms quadratic in $U$. Two of these terms are 
due to the localized spin fluctuations (LSF), 
represented by the spin-susceptibility bubble $({\cal G}{\bar {\cal G}})$, 
mutually interacting at the impurity site and also with the induced
localized pairing fluctuations (LPF), shown as the pairing-susceptibility
bubble  $({\cal F}{\bar {\cal F}})$. The third contribution arises
from the induced mutually interacting localized pairing fluctuations. 

Preliminary numerical results [13] indicate the doubling of 
bound states, in comparison to the HF theory. In particular, 
the new bound states tend towards $\omega =0$ for increasing $U$. 
We shall discuss the full numerical results in detail elsewhere.  
Our approach can also be readily extended to other situations 
of interest, such as an Anderson impurity in unconventional 
superconductors [14]. In this case, the self-energy expressions 
are formally the same but the order parameter
must in general be interpreted as a matrix in spin space. Also the order 
parameter $\Delta (\hat k)$ for unconventional superconductors
possesses less rotational symmetry than that in the $s$-wave 
case. This will naturally lead to a $\hat k$-dependence of the 
$d$-electron Green's function $\hat G_S(\hat k, \omega )$, 
the matrix selfenergy $\hat \Sigma _S(\hat k, \omega )$ and 
the bound-state spectrum below the $\hat k$-dependent 
energy-gap edge. 

\newpage
\section*{Acknowledgments}

We are grateful to Professors G.\ Sch\"on, H.\ Shiba and P.\ W\"olfle 
for helpful discussions. This work has been supported by the Academy of 
Finland and the Swiss National Science Foundation; we thank Professors 
G.\ Blatter and T.\ M.\ Rice for cordial hospitality at ETH Z\" urich where
part of this work was performed. This work is supported by the NSF under 
Grant No.\ DMR-9704972 and under the Materials Research Science and 
Engineering Centre Program, Grant No.\ DMR-96-32527. One of us (ATA) 
is grateful to Helsinki University of Technology for a postgraduate research 
award; another one (MMS) thanks the Department of Physics at the 
University of Wisconsin-Madison for cordial hospitality during 
the preparation of this paper.

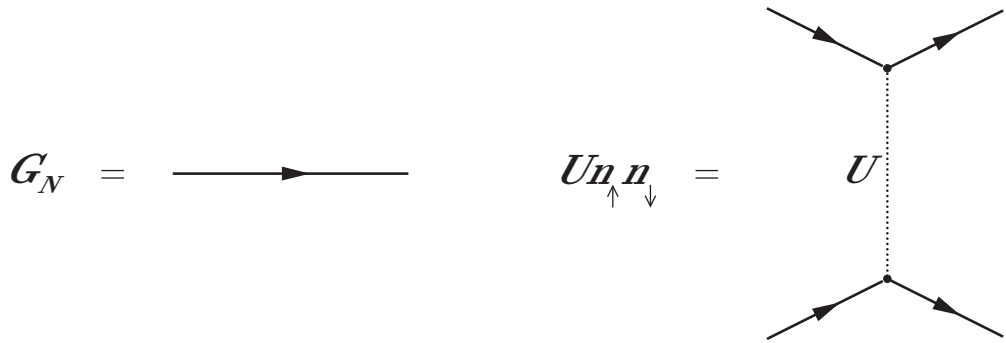
\begin{figure}
\caption{The local $d$-electron propagator in the normal state, 
$G_N(\omega)$, is here denoted as a line with an arrowhead (l.h.s.). 
The local Coulomb-repulsion or Anderson-Hubbard interaction term
$Un_{\uparrow }n_{\downarrow }$ is represented in the diagrams 
with a dashed line having two vertices (r.h.s.).}
\label{interaction}
\end{figure} 

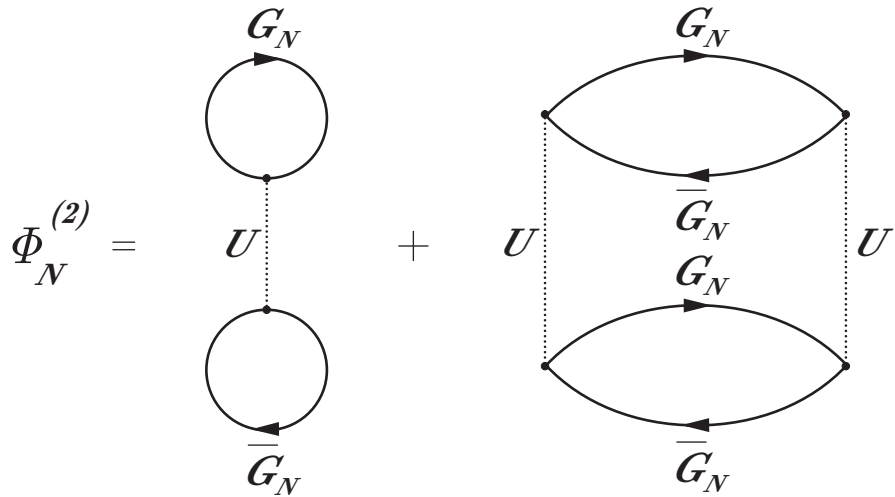
\begin{figure}
\caption{The terms $\Phi ^{(2)}_N$ in the normal state up to the second 
order in $U$. The linear term in $U$ is the Hartree-Fock bubble.
The second-order term in $U$ contains two particle-hole (susceptibility) bubbles, describing interacting localized spin fluctuations (LSF) at the impurity site.}
\label{normal}
\end{figure}

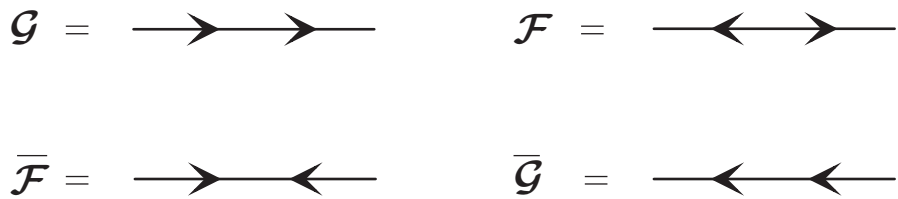
\begin{figure}
\caption{Elements of the localized $2\times 2$ Green's-function matrix 
in the Nambu space for the superconducting state are represented by lines 
with double arrows. Here ${\cal G}(\omega )$, corresponding to the 
$G_N(\omega )$ in Fig.~1, is the electron propagator, while 
${\bar {\cal G}}(\omega )$ is the time-reversed hole-propagator function; 
${\cal F}(\omega )$ is the anomalous particle-hole Green's function 
and ${\bar {\cal F}}(\omega )$ denotes its conjugate.}
\label{nambu}
\end{figure}

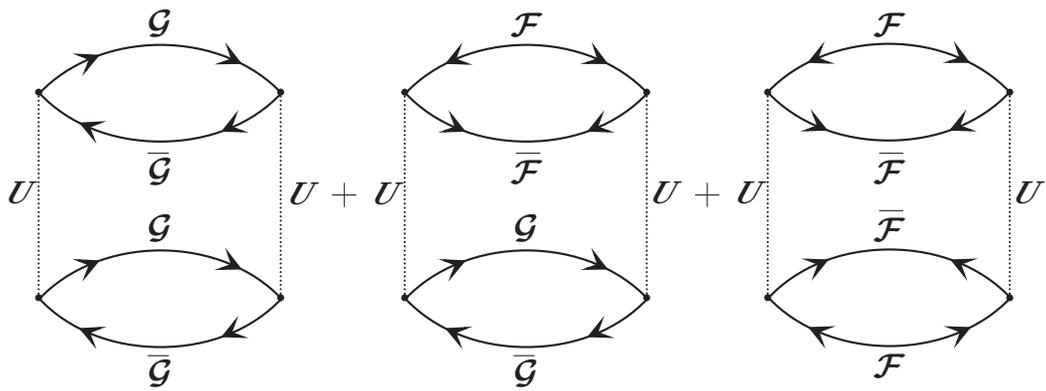
\begin{figure}
\caption{Generalization of $\Phi$ to the second order in $U$ for the 
pair-correlated state, $\Phi ^{(2)}_S$, expressed in terms of the 
superconducting propagators, ${\cal G}$, ${\bar {\cal G}}$, ${\cal F}$, 
and ${\bar {\cal F}}$, in Fig.~3. Due to the anomalous propagators 
in the pair-correlated medium, there now occur two first-order 
Hartree-Fock terms linear in $U$ and three second-order terms in $U$, 
owing to the interacting localized spin fluctuations (LSF), correlated localized
spin and pairing fluctuations (LSPF), and interacting localized  
pairing fluctuations (LPF), respectively.}
\label{super}
\end{figure}

\end{document}